\documentclass[a4paper]{iopart}
\usepackage{graphicx}
\usepackage{amsfonts}
\usepackage{iopams}

\expandafter\let\csname equation*\endcsname\relax
\expandafter\let\csname endequation*\endcsname\relax
\usepackage{amsmath}
\usepackage{amsthm}
\usepackage{amssymb}

\begin{document}
\title{Classical-quantum correspondence for shape- invariant systems}

\author{A.M. Grundland $^{1,2}$ , D. Riglioni $^1$}
 
\address{ $^1$ Centre de Recherches Math\'ematiques, Universit\'e de Montr\'eal, C.P. 6128 Succ. Centre-ville Montreal (QC) H3C3J7, Canada
\newline $^2$ Department of Mathematics and Computer Science, Universit\'e du Qu\'ebec, Trois-Rivi\`eres, C.P. 500 (QC) G9A5H7, Canada}

\ead{grundlan@crm.umontreal.ca} 

\ead{riglioni@crm.umontreal.ca}

\begin{abstract} 
A quantization procedure, which has recently been introduced for the analysis of Painlev\'e equations, is applied to a general time-independent potential of a Newton equation. This analysis shows that the quantization procedure preserves the exact solvability property for the class of shape-invariant potentials.  When a general potential is considered the quantization procedure involves the solution of a Gambier XXVII transcendental equation. Explicit examples involving classical and exceptional orthogonal Laguerre and Jacobi polynomials are discussed. 
\newline \phantom{.}
\end{abstract}



\section{Introduction}

The linearization of a nonlinear differential equation $\Delta(x,t,[u])$ \\ (where  $[u] \equiv (u, u_t,u_x,u_{tt},...)$) as the compatibility condition of an overdetermined system (i.e. the zero curvature condition (ZCC)) of the linear differential equations

\begin{equation}
\label{start}
\begin{cases}
\partial_x \Phi = {\bf U} \Phi ,\\
\partial_t \Phi = {\bf V} \Phi ,
\end{cases} \rightarrow \partial_t {\bf U} - \partial_x {\bf V} + [{\bf U} , {\bf V}] \equiv \Delta(x,t,[u(x,t)]) = 0 ,
\end{equation}

\noindent is a well-known technique which, over the years, has allowed for a systematic investigation of many important integrable nonlinear partial differential equations (PDEs) such as the KdV, nonlinear Schroedinger and Sine-Gordon equations (see e.g. \cite{calogero} \cite{ablovitz}). This technique can also be used to study nonlinear ordinary differential equations (ODEs). In this case the second independent variable, say the $x$ of (\ref{start}), is replaced by the spectral parameter $\lambda$. A remarkable example in this class of nonlinear ODEs is represented by the Painlev\'e equations \cite{bobenko}. Painlev\'e equations arise in many contexts and they can also be defined as particular reductions of some integrable PDEs solvable by the inverse scattering transfom (see e.g. \cite{cit4} - \cite{painleve2}). 
The idea of regarding each Painlev\'e equation as the compatibility condition of a set of linear differential equations goes back to the work of R. Fuchs \cite{fuchs}.
In some recent papers by Suleimanov \cite{suleimanov} and by Zabrodin and Zotov \cite{zabrodinzotov}-\cite{zabrodinzotov3} such a mechanism has been used to define the Painlev\'e equations in the so-called "Calogero form". The study of Painlev\'e equations as a Hamiltonian system (also known as "Calogero form") has a long history (for some relevant references see e.g. \cite{malmquist} \cite{okamoto} and references therein). In particular, in \cite{levin} it is shown that for all Painlev\'e equations $P(\ddot{y}, \dot{y},y,t) = 0$  it is possible to find a transformation $(y,T)\rightarrow (u,t), \quad y=y(u,t),T=T(t)$ which maps the Painlev\'e equation to the Newton differential equation  

\begin{equation}
\label{newton1}
\ddot{u} = -\partial_u V_c(u,t),
\end{equation}

\noindent (where the dot denotes the derivative with respect to $t$). Equation (\ref{newton1})  can be regarded as the equation of motion for a time-dependent Hamiltonian system 

\begin{equation}
\label{classicham}
H(p,u,t) = \frac{p^2}{2} + V_c(u,t). 
\end{equation}

\noindent The main result of Zabrodin and Zotov \cite{zabrodinzotov} is the fact that equation (\ref{newton1}) can be regarded as the compatibility condition of a linear spectral problem (LSP) which turns out to be mathematically equivalent to a time-dependent Schroedinger equation 

\begin{equation}
\label{schroedinger1}
-i \partial_t \psi(x,t) = -\frac{1}{2} \partial_x^2 \psi + V_q(x,t) \psi , 
\end{equation}

\noindent where $x$ plays the role of the spectral parameter and the quantum potential $V_q(\cdot,t)$ turns out to be identical to the potential $V_c(\cdot,t)$ up to some renormalization in the parameters contained in $V_q(\cdot,t)$. For this reason, we will refer to equation (\ref{schroedinger1}) as the quantization of  equation (\ref{classicham}) in the sense of the classical-quantum correspondence as introduced by Suleimanov, Zabrodin and Zotov (SZZ) (see \cite{suleimanov} \cite{zabrodinzotov}). For this paper to be self-contained, let us explicitly recall the notion of the classical-quantum correspondence for the Painlev\'e equations PIV and PV. 

 We first consider the LSP for Painlev\'e equation PIV \cite{zabrodinzotov}

\begin{eqnarray}
\nonumber
\partial_x \left( \begin{array}{c} \phi_1 \\ \phi_2 \end{array} \right) & = & \left( \begin{array}{cc} \frac{x^3}{2} +tx + \frac{Q+\frac{1}{2}}{x} & x^2 -u^2 \\ \frac{Q^2+\frac{\beta}{2}}{u^2x^2} -Q-\alpha-1 & - \frac{x^3}{2} -tx - \frac{Q+\frac{1}{2}}{x} \end{array}\right) \left( \begin{array}{c} \phi_1 \\ \phi_2 \end{array} \right) ,\\ \label{lsp1}
\\
\nonumber
\partial_t \left( \begin{array}{c} \phi_1 \\ \phi_2 \end{array} \right)  & = & \left( \begin{array}{cc} \frac{x^2+u^2}{2} + t & x \\  -\frac{Q + \alpha + 1}{x} & -\frac{x^2+u^2}{2}-t \end{array}\right) \left( \begin{array}{c} \phi_1 \\ \phi_2 \end{array} \right) ,
\end{eqnarray}
\noindent where
\begin{eqnarray}
Q & = & u \dot{u} - \frac{u^4}{2} - t u^2. \nonumber
\end{eqnarray}

\noindent The compatibility condition associated with the LSP (\ref{lsp1}) corresponds to the Painlev\'e PIV equation in the Calogero form \cite{zabrodinzotov}
\begin{equation}
\label{newtonPIV}
\Delta_{PIV} = \ddot{u} + \partial_u Vc = 0 ,
\end{equation}

\begin{equation}
\label{VPIV}
V_c = -\frac{u^6}{8} - \frac{t u^4}{2} -\frac{1}{2}(t^2-\alpha) u^2 + \frac{\beta}{4 u^2}.
\end{equation}

\noindent Moreover, from the LSP (\ref{lsp1}) we observe that the function $\psi = e^{\int^t (\frac{\dot{u}^2}{2} + V_c)dt'} \phi_1$ satisfies the following  non-stationary "real" Schroedinger equation (\ref{schroedinger1})

\begin{equation}
\label{schroedingerPIV}
\partial_t \psi = \frac{1}{2} \partial_x^2 \psi + \left( -\frac{x^6}{8} - \frac{t x^4}{2} -\frac{1}{2}(t^2-\alpha) x^2 + \frac{\beta + \frac{1}{2}}{4 x^2} \right) \psi ,
\end{equation}

\noindent where the potential in the Schroedinger equation (\ref{schroedingerPIV}) turns out to be the same as that of the classical equation (\ref{newtonPIV}) up to a shift in the parameter $\beta$. An analogous analysis can be performed for the Painlev\'e equation PV. 

Let us introduce the following LSP

\begin{equation}
\label{eq9lsp}
\partial_x \left( \begin{array}{c} \phi_1 \\ \phi_2 \end{array} \right)  =  {\bf U} \left( \begin{array}{c} \phi_1 \\ \phi_2 \end{array} \right) \quad , \quad  \partial_t \left( \begin{array}{c} \phi_1 \\ \phi_2 \end{array} \right) =  {\bf V} \left( \begin{array}{c} \phi_1 \\ \phi_2 \end{array} \right), 
\end{equation}

where the entries of ${\bf U}, {\bf V}$ are  given by 

\begin{eqnarray}
{\bf U}_{11} & = & \dot{u} \frac{\sinh 2u}{\sinh 2x} - \frac{2 \sigma}{\sinh 2x} \left( \cosh 2x - \cosh 2u \right) + \\
 & & + \frac{e^{2t}}{4 \sinh 2x} \left( \cosh 4x - \cosh 4u \right) + \coth (2x) , \nonumber \\ 
{\bf U}_{12} & = & e^t \left( \cosh 2x - \cosh 2u \right), \\
{\bf U }_{21} & = & \dot{u}^2 \frac{e^{-t}}{\sinh^2 2x} \left( \cosh 2u + \cosh 2x \right) + \\
 & & \dot{u} \frac{\sinh 2u}{\sinh^2 2x} \left( 4 \sigma e^{-t} - e^t (\cosh 2u + \cosh 2x ) \right) + \nonumber \\
& & 8 \sigma^2 e^{-t} \frac{\coth^2 u}{\sinh^2 2x} \left( \sinh^2 u - \cosh^2 x \right) - 2 \sigma e^t \frac{\sinh^2 2u}{\sinh^2 2x} + \nonumber \\
& & -2 e^{-t} \frac{\xi^2 + 2 \xi \sigma}{\sinh^2 u \sinh^2 x} + \frac{2e^{-t} \zeta^2}{\cosh^2 u \cosh^2 x} + \nonumber \\ 
& & \frac{e^{3t} \sinh^2 2u}{4 \sinh^2 2x} \left( \cosh 2u + \cosh 2x \right) , \nonumber \\
{\bf U}_{22} & = & - {\bf U}_{11}, 
\end{eqnarray}

\noindent and

\begin{eqnarray}
{\bf V}_{11} \! & = & \frac{1}{2} e^{2t} \left( \cosh 2x + \cosh 2u \right) -2 \sigma + \frac{1}{2}, \\
{\bf V}_{12} \! & = & e^t \sinh 2x , \\
{\bf V}_{21} \! & = & \! \frac{e^{-t}}{\sinh \! 2x} \! \left( \!\! \left( \! \dot{u} \! - \! \frac{e^{2t}\! \sinh 2 u}{2}   \! \right)^2 \!\!\!\! + \! \frac{4 \zeta^2}{\cosh^2 \! u} \! - \! \frac{4 \xi^2 \! + \! 8 \xi \sigma}{\sinh^2 \! u} \! - \! 4 \sigma^2 \! \coth^2 \! u \!\! \right) \!\!, \\
{\bf V}_{22} \! & = & - {\bf V}_{11} , 
\end{eqnarray} 

\noindent where $\sigma, \xi, \zeta$ are free parameters. The compatibility condition reduces to the following Newton equation for the Painleve equation PV 

\begin{eqnarray}
\label{newtonpv}
\Delta_{PV} \! & = & \ddot{u} + \partial_u V_c = 0 ,\\
\label{VPV} 
V_c \! & = & \frac{4 \zeta^2}{2 \cosh^2 \! u} \! - \!  \frac{4(\xi \! + \! \sigma)^2}{2 \sinh^2 \! u} \! - \! \frac{e^{4t}}{16} \cosh 4u \! + \! \left( \! \sigma \! - \! \frac{1}{2} \! \right) \! e^{2t} \! \cosh 2u  . 
\end{eqnarray}

\noindent The function $\phi_1$  satisfies a "real" Schroedinger equation for the function \\ 

\begin{equation}
\psi = \phi_1 e^{\int^t (\frac{\dot{u}^2}{2} + V_c)dt'} 
\end{equation}

\begin{equation}
\label{schroedingerPV}
\partial_t \psi = \frac{1}{2} \partial_x^2 \psi + \left(  \frac{4 \zeta^2 - \frac{1}{4}}{2 \cosh^2 x} - \frac{4(\xi + \sigma)^2 - \frac{1}{4}}{2 \sinh^2 x} - \frac{e^{4t}}{16} \cosh 4x + \left( \sigma - \frac{1}{2} \right) e^{2t} \cosh 2x  \right) \psi ,
\end{equation}

\noindent which corresponds to the quantization of equation (\ref{newtonpv}) up to a redefinition of the parameters $\zeta,\xi$ and $\sigma$.

\noindent We remark that both the Painlev\'e PIV and Painlev\'e PV equations  can be regarded as time-dependent integrable deformations for the potential of a harmonic oscillator with centrifugal barrier and the Poschl Teller potential, respectively

\begin{eqnarray}
\label{VPIV}
V_{PIV} & \rightarrow & \frac{1}{2} \alpha x^2 + \frac{\beta}{x^2}, \\
\label{VPV}
V_{PV} & \rightarrow & \frac{4 \zeta^2}{2 \cosh^2 x} - \frac{4(\xi + \sigma)^2}{2 \sinh^2 x}. 
\end{eqnarray}  
These potentials (\ref{VPIV}) and (\ref{VPV}) are well-known for being "shape-invariant". Shape- invariant potentials were implicitly introduced  by Schroedinger in \cite{schroedinger} and then generalized by Infeld and Hull in \cite{infeld}  as a mechanism to solve algebraically the bounded spectrum of a quantum mechanical system (for a more recent review on the topic see e.g. \cite{cooper}). 

Let us briefly recall the definition of a shape-invariant potential as a potential whose Hamiltonian operator can be factorized through two ladder operators $a_{{\bf \lambda}}, a^\dagger_{{\bf \lambda}}$ 

\begin{eqnarray}
\hat{H} & = & a^\dagger_{{\bf \lambda}} a_{{\bf \lambda}} = -\partial_x^2 + V_{{\bf \lambda}}(x),\\
a_{{\bf \lambda}} & = & -i \partial_x +i W'_{{\bf \lambda}}(x) , \\
a^\dagger_{{\bf \lambda}} & = &    -i \partial_x - i W'_{{\bf \lambda}}(x) ,
\end{eqnarray} 

\noindent  having the following property 

\begin{equation}
\label{shapeinvariant}
a_{{\bf \lambda}} a^\dagger_{{\bf \lambda}} = a^\dagger_{{\bf \lambda} + {\bf \delta}} a_{{\bf \lambda} + {\bf \delta}} + const ,
\end{equation}

\noindent where ${\bf \lambda}, {\bf \delta}$ are in general parameter vectors. It is straightforward to verify by direct computation that property (\ref{shapeinvariant}) holds for the potentials (\ref{VPIV}) and (\ref{VPV}), if the ladder operators take the form 

\begin{eqnarray}
a_l & = & -i \partial_x + i(-\omega x + \frac{l}{x}), \\
a^\dagger_l a_l & = & -\partial_x^2 + \frac{l(l-1)}{x^2} + \omega^2 x^2 -2 \omega l - \omega ,\\
a_l a^\dagger_l & = & -\partial_x^2 + \frac{l(l+1)}{x^2} + \omega^2 x^2 -2 \omega l + \omega ,
\end{eqnarray}

\noindent and

\begin{eqnarray}
a_{l,g} & = & -i \partial_x + i(g \coth x + l  \tanh x ) , \\
a^\dagger_{l,g} a_{l,g} & = & -\partial_x^2 + \frac{g(g-1)}{\sinh^2 x} - \frac{l(l-1)}{\cosh^2 x} + (g+l)^2 , \\
a_{l,g} a^\dagger_{l,g} & = &-\partial_x^2 + \frac{g(g+1)}{\sinh^2 x} - \frac{l(l+1)}{\cosh^2 x}+ (g+l)^2 . 
\end{eqnarray}

On the basis of the above considerations the principal objective of this paper is to show that whenever the potential $V_c(u(t),t)$ of the Newton equation (\ref{newton1}) does not depend explicitly on time it is possible to define a LSP whose compatibility condition involves the solution of the equation (\ref{newton1}) and of a nonlinear differential equation which can be reduced (under some specific assumptions) to the Gambier equation XXVII (GXXVII). In particular we will show that the exact solvability of GXXVII is connected with the exact solvability of the "quantization" of (\ref{newton1}). In fact it turns out that if the potential $V$ has the shape-invariant property then it is possible to provide an exact solution of GXXVII in terms of orthogonal polynomials either classical or exceptional.

The present paper is organized as follows. In section 2, we recall the basic concepts necessary for an understanding of the quantization in the sense of the SZZ. In particular we will provide the master equation which allows us to connect any Newton equation (\ref{newton1}) to its corresponding Schroedinger equation (\ref{schroedinger1}). Particular solutions will be provided in subsections 2.1 and 2.2 for quantum Schroedinger equations characterized by potentials which are shape-invariant. In section 3 we will discuss in detail the classical quantum correspondence for the harmonic oscillator system with centrifugal barrier providing the exact solution of the LSP in terms of exceptional orthogonal Laguerre and Jacobi polynomials. Section 4 contains final remarks and possible future developments.

\section{Linear spectral problem and nonstationary Schroedinger equation}
We start by considering a completely general LSP defined by two potential matrices $\tilde{U},\tilde{V} \in \mathfrak{sl}(2)$. 

\begin{equation}
\begin{cases}
\partial_x \tilde{\Phi} = \tilde{\mathbf{U}} \tilde{\Phi}, \\
\partial_t \tilde{\Phi} = \tilde{\mathbf{V}} \tilde{\Phi},
\end{cases}
\end{equation} 
 
\noindent where $\tilde{U},\tilde{V}$ are given by the traceless matrices 

\begin{equation}
\tilde{\mathbf{U}} = 
\left(
\begin{array}{cc}
a & b \\
c & -a
\end{array}
\right), \quad \tilde{\mathbf{V}} = 
\left(
\begin{array}{cc}
\tilde{A} & B \\
C & -\tilde{A}
\end{array}
\right).
\end{equation} 

\noindent We reduce the number of undetermined functions in the entries of matrices $\tilde{\bf U}$ and $\tilde{\bf V}$ by considering the following gauge transformation

\begin{equation}
\tilde{\Phi} = {\mathbf{T}} \Phi , \quad {\mathbf{T}} = 
\left(
\begin{array}{cc}
1 & 0 \\
-a / b & 1
\end{array}
\right)  ,
\end{equation} 

\begin{eqnarray}
{U} &=& {\mathbf{T}}^{-1} \tilde{\mathbf{U}} {\mathbf{T}} - {\mathbf{T}}^{-1} \partial_x {\mathbf{T}} =  \left(
\begin{array}{cc}
0 & b \\
\alpha & 0
\end{array}
\right)  , \\
{V} &=& {\mathbf{T}}^{-1} \tilde{\mathbf{V}} {\mathbf{T}} - {\mathbf{T}}^{-1} \partial_t {\mathbf{T}} =\left(
\begin{array}{cc}
{A} & B \\
\beta & -{A}
\end{array}
\right) ,
\end{eqnarray}

\noindent where $A$, $\alpha$ and $\beta$ are given by

\begin{eqnarray}
{A} = \tilde{A} - \frac{a B}{b}, \\
\alpha = \frac{1}{b} (-det(\tilde{{\mathbf{U}}}) + a_x) - \frac{a b_x}{b^2}, \\
\beta = \frac{a^2 B}{b^2} + \frac{b C + a_t}{b} + \frac{a}{b^2} (2 A b - b_t).
\end{eqnarray}
  
\noindent The LSP

\begin{equation}
\label{eq2}
\begin{cases}
\partial_x \Phi = {\mathbf{U}} \Phi ,\\
\partial_t {\Phi} = {\mathbf{V}} {\Phi} , 
\end{cases} \quad \Phi = (\phi_{ij}), \quad i,j = \{ 1,2 \}
\end{equation} 

\noindent can be rewritten as follows

\begin{eqnarray}
\label{eq1}
{\Phi_{1,i}}_x &=& b \Phi_{2,i} \rightarrow {\Phi_{1,i}}_{xx} = b_x \Phi_{2,i} + b {\Phi_{2,i}}_x ,\\
{\Phi_{2,i}}_x &=& \alpha \Phi_{1,i},
\end{eqnarray}

\noindent where $\alpha, \beta$ are functions to be determined. The equation (\ref{eq1}) defines a Sturm Liouville problem if we replace 

\begin{eqnarray}
\Phi_{2i} = \frac{{\Phi_{1i}}_x}{b}, \\
{\Phi_{2i}}_x = \alpha \Phi_{1i}.
\end{eqnarray}

\noindent In addition, the entry $\Phi_{1,i}$ has to satisfy another linear PDE which can be defined using the second equation in the LSP (\ref{eq2}) 

\begin{equation}
\label{eqq3}
\partial_t \Phi = { \bf{V}} \Phi \rightarrow {\Phi_{1,i}}_t = A \Phi_{1,i} + B \Phi_{2,i} \rightarrow \phi_{2,i} = \frac{1}{B} {\Phi_{1,i}}_t - \frac{A}{B} \Phi_{1,i}.
\end{equation}

\noindent Replacing (\ref{eqq3}) in (\ref{eq1}) we obtain

\begin{equation}
{\Phi_{1,i}}_{xx}  = \frac{b_x}{B} {\Phi_{1,i}}_t + (b\alpha - \frac{b_x A}{B} ) {\Phi_{1,i}} .
\end{equation}

\noindent This equation turns into the Schroedinger equation (\ref{schroedinger1}) if we set 

\begin{equation}
\begin{cases}
B = \frac{i b_x}{2} , \\
\alpha = \frac{2}{b} (V_q(x) - i A) ,
\end{cases}
\end{equation}

\begin{equation}
i {\Phi_{1,i}}_{t} = -\frac{1}{2} {\Phi_{1,i}}_{xx} +V_q(x) \Phi_{1,i} .
\end{equation}

\noindent The final step consists in making the system  (\ref{eq2}) a compatible system, namely to impose the ZCC. 

\begin{equation}
ZCC \equiv \partial_t {\bf U} - \partial_x {\bf V} + [{\bf U} , {\bf V}] = 0 .
\end{equation}

\noindent From the components of $ZCC_{1,2}$ and $ZCC_{1,1}=ZCC_{2,2}$ we can determine the functions $A$ and $\beta$ 

\begin{equation}
\begin{cases}
A = \frac{1}{2b} \left( b_t - \frac{i}{2}b_{xx} \right) ,\\
\beta = \frac{1}{b} \left( A_x + \frac{i}{2} \alpha b_x \right).
\end{cases}
\end{equation}

\noindent The last condition $ZCC_{2,1}=0$ can be used to fix the function $b(x,t)$. However, since the goal is to connect the quantum linear problem to classical mechanics we require that the function $b(x,t)$ be dependent on time through the time-dependent variable $u(t)$ , $b = b(x,u(t))$. Moreover we set the function $u(t)$ such that it satisfies the conservation of energy  
for a classical system 
\begin{equation}
\frac{\dot{u}^2}{2} + V_c(u(t)) = 0.
\end{equation} 

\noindent With this assumption $ZCC_{2,1}=0$ turns into the following PDE

\begin{equation}
\label{mastereq}
4 i \dot{u} (b b_{xxu} \!-\! b_x b_{xu}) \!-\! 8 V_c b_u^2 \!+\! 8 V_q b_x^2 \!+\! b_{xx}^2  \!-\! 2b_x b_{xxx} \!+\! b (4 V_c' b_u \!+\! 8 V_c b_{uu} \!-\! 4 V_q' b_x \!-\! 8 V_q b_{xx} \!+\! b_{xxxx}) = 0,
\end{equation}

\noindent where we have denoted $V_c' \equiv V_{c,u}$ and $V_q' \equiv V_{q,x}$. Considering our purposes we can simplify the nonlinear PDE (\ref{mastereq}) with non-constant coefficients to two nonlinear ODEs by assuming the following ansatz on the function $b(x,u(t))$ 

\begin{equation}
\label{eq52}
b = b_1(x) - b_2(u(t)).
\end{equation}

\noindent In particular equation (\ref{mastereq}) turns out to be satisfied whenever the two ODEs 

\begin{equation}
\label{vc}
V_c {b_2}_u^2 = k_1 b_2^2 + k_2 b_2 + k_3 , \quad k_1,k_2,k_3 \in \mathbb{Re} ,
\end{equation}

\begin{equation}
\label{vq}
V_q {b_1}_x^2 = k_1 b_1^2 + k_2 b_1 + k_3 + \frac{{b_1}_x {b_1}_{xxx}}{4} - \frac{{b_1}_{xx}^2}{8},
\end{equation}

\noindent hold. The function $b_2$ can easily be determined for any potential $V_c$  by integrating the first-order elliptic ODE (\ref{vc})

\begin{equation}
\int \frac{d b_2}{\sqrt{k_1 b_2^2 + k_2 b_2 + k_3}} = \int \frac{1}{\sqrt{V_c(u)}} du.
\end{equation}

\noindent On the other hand we have to solve a nonlinear third-order ODE (\ref{vq}) in order to obtain the function $b_1(x)$ which produces a given quantum potential $V_q(x)$. If we set the free parameters $k_2=k_3=0$ we can reduce (\ref{vq}) to the second order ODE

\begin{equation}
\label{gambier27}
b_1(x) = e^{\int^x f(s) ds} \rightarrow f f_{xx} +4k_1 - 4 V_q(x) f^2 +2f^2f_x - \frac{f_x^2}{2} + \frac{f^4}{2} = 0.
\end{equation}

\noindent The equation (\ref{gambier27}) turns out to be the Gambier equation GXXVII  (\cite{gambier}). Such an equation can be linearized to a linear 4th-order ODE. However,  it is possible to express (\ref{vq}) through a system of two Sturm-Liouville problems
for the functions $\rho$ and $b_1$

\begin{equation}
\label{sch2}
\begin{cases}
\rho_{xx} = ( 2 V_q - \epsilon \sqrt{2k_1} ) \rho ,\\
{b_1}_{xx} = 2 \frac{\rho_x}{\rho} {b_1}_x + 2 \epsilon  \sqrt{2k_1} b_1, \quad \epsilon^2 =1,
\end{cases}
\end{equation}

\noindent which is equivalent to the 4th order ODE 

\begin{equation}
{V_q}_x {b_1}_x + 2 V_q {b_1}_{xx} = 2 k_1 b_1 + \frac{{b_1}_{xxxx}}{4}.
\end{equation}

\noindent The parameter $k_2$ can be introduced with the shift $b \rightarrow b + \frac{k_2}{k_1}$ and $k_3$ is a constant of integration.

In particular if we assume that the solution for $\rho$ in (\ref{sch2}) can be expressed as $\rho = e^{W(x)}$ with prepotential $W(x)$, then the quantum potential $V_q$ can be expressed as $2 V_q = \epsilon \sqrt{2k_1} + W_{xx} + W_x^2$ . Under these assumptions we can recast the Sturm Liouville problem involving $b_1$ as a Schroedinger equation by introducing the following gauge transformation $b_1 = e^{W} \psi $

\begin{equation}
\label{eq59}
\psi_{xx} = (W_x^2 -W_{xx} + 2 \epsilon \sqrt{2k_1}) \psi .
\end{equation}

\noindent The solution of (\ref{vq}) reduces to the solution of a couple of Schroedinger equations (\ref{sch2}) involving $V_q$ as a potential. In order to provide some explicit example, let us consider the potential $V_q$ characterized by the shape-invariant property.

\begin{equation}
\label{susyzero}
\begin{cases}
a = -\partial_x + W_x , \\
a^{\dagger} = \partial_x + W_x ,
\end{cases}
\end{equation}

\noindent such that 

\begin{equation}
\label{prepoteq}
W_x^2 + W_{xx} = 2 V_q - \epsilon \sqrt{2k_1}. 
\end{equation}

\noindent The system (\ref{sch2}) turns into the Schroedinger equations

\begin{equation}
\label{susy}
\begin{cases}
a^\dagger a \rho = 0 , \\
(a a^\dagger + 2 \epsilon \sqrt{2k_1} ) \psi = 0 .
\end{cases}
\end{equation}

\noindent  It is straightforward to verify that, given a basis of eigenfunctions for the operator $a^\dagger a \phi = \lambda \phi $, we obtain 

$$\psi = a \phi ,$$ 

\noindent where $k_1$ depends on the eigenvalue $\lambda$ of the eq (\ref{eq59}). 

Now we present some examples in order to illustrate the above theoretical considerations.

\subsection{{ \bf The harmonic oscillator potential} }
As a simple example we consider the harmonic oscillator potential

\begin{equation}
\label{oscillatore}
2 V_q + \sqrt{2k_1} = \omega^2 x^2 + \frac{l(l-1)}{x^2} + \omega(2l+1+4N).
\end{equation}

\noindent Such a quantum potential $V_q$ can be obtained by setting the prepotential $W$ as follows

\begin{equation}
W = \frac{\omega}{2} x^2 + l \ln (x) + \ln (L_N^{l-\frac{1}{2}}(-\omega x^2)) ,
\end{equation}

\noindent where  $L_N^l$ are the Laguerre polynomials.
The potential $V_q$ in  (\ref{oscillatore}) doesn't depend on the sign of $\omega$. We assume from now on that $\omega$ is a positive real number in order to avoid singularites in the Schroedinger equations (\ref{susy}). Under these assumptions it is easy to determine the function $\psi$ from the eigenfunctions of the Harmonic oscillator

\begin{equation}
\phi_{n,l} = e^{-\frac{\omega x^2}{2}} x^l L_n^{l-\frac{1}{2}}(\omega x^2),
\end{equation}
\noindent which satisfy the equation

\begin{equation}
\label{eq66}
a^\dagger a \quad \!\! \phi_{n,l} = \omega (4l +2 +4 N +4n) \phi_{n,l} .
\end{equation}

\noindent From equation (\ref{eq66}) we obtain that $\psi = a \phi_{n,l}$ satisfies 

\begin{equation}
(a a^\dagger -2\sqrt{2 k_1})\psi =0 ,\quad k_1 = \frac{\omega^2}{2}(2l + 1 + 2 N + 2 n)^2  ,
\end{equation}

\noindent and the quantum potential $V_q$ turns out to be

\begin{equation}
V_q = \frac{\omega^2}{2} x^2 + \frac{l(l-1)}{2x^2} + \omega(N-n) .
\end{equation}

\noindent Also, the function $b_1$ takes the form 

\begin{equation}
b_1 = e^W \psi = 2 \omega x^{2l+1} P(N,n,l,\omega x^2),
\end{equation}

\begin{equation}
P(N,n,l,\omega x^2) = L_N^{l + \frac{1}{2}} (-\omega x^2) L_n^{l + \frac{1}{2}} (\omega x^2) - L_{N-1}^{l + \frac{1}{2}} (-\omega x^2) L_{n-1}^{l + \frac{1}{2}} (\omega x^2),
\end{equation}

\noindent where $P(N,n,l,\omega x^2)$ are the exceptional Laguerre orthogonal polynomials. Exceptional orthogonal polynomials have been introduced quite recently by Gomez-Ullate et al. in \cite{gomez}. The introduction of these new orthogonal polynomials led to new families of shape-invariant potentials (see e.g. the papers of Quesne \cite{quesne} and Sasaki \cite{sasaki}).

\subsection{ { \bf The Poschl-Teller potential } }

Another interesting example is provided by the solution of the ODE (\ref{vq}) when the potential $V_q$ is given by the Poschl-Teller 

\begin{equation}
2 V_q + \sqrt{2k_1} = \frac{(g+N)(g+N+1)}{\sin^2 x} + \frac{(h+N-1)(h+N-2)}{\cos^2 x} - (2N+h-g-1)^2 .
\end{equation}

\noindent In this case the prepotential $W$ takes the form 

\begin{equation}
W = -(g+N) \ln (\sin (x)) + (h+N-1) \ln (\cos (x)) + \ln (P_N^{-g-N-\frac{1}{2},h+N-\frac{3}{2}}(\cos(2x))) ,
\end{equation}

\noindent where $P_N^{\alpha,\beta}$ are the Jacobi polynomials. We introduce the following set of eigenfunctions for the Hamiltonian operator defined by $a^\dagger a$ 

\begin{equation}
\label{eq73}
\phi_{n,g+N,h}(x) = \sin^{g+N+1}(x) \cos^{h+N-1}(x) P_{n}^{g+N+\frac{1}{2},h+N-\frac{3}{2}} (\cos(2x)) ,
\end{equation}

\noindent which satisfies the following equation

\begin{equation}
a^\dagger a \phi_{n,g+N,h} = (2n+1+2g)(4N+2n+2h-1) \phi_{n,g+N,h} .
\end{equation}

\noindent From (\ref{eq73}) we obtain the wavefunction in the form $\psi =  a \phi_{n,g+N,h}$  which satisfies 

\begin{equation}
(a a^\dagger -2 \sqrt{2k_1}) \psi =0; k_1 = \frac{1}{8} (2n +1 +2g)^2 (2n+2h+4N-1)^2 ,
\end{equation}

\noindent and the potential $V_q$ turns out to be 

\begin{equation}
V_q = \frac{(g \! + \! N)(g \! + \! N \! + \! 1)}{2 \sin^2 \! x} + \frac{(h \! + \! N \! - \! 1)(h \! + \! N \! - \! 2)}{ 2 \cos^2 x} - \frac{(2N \! + \! h \! - \! g \! - \! 1)^2 + (2n \! + \! g \! + \! h \! + \! 2N)^2}{4} . 
\end{equation}

\noindent As previously noted the function $b_1$ turns out to be $b_1 = \psi e^W $.
In this case the function $b$ given by (\ref{eq52}), can be expressed in terms of exceptional orthogonal polynomials

\begin{equation}
e^W \psi = -(2n+1+2g) \cos(x)^{2h+2N-1} Pe_{n,N}^{g,h}(\cos(2x)) ,
\end{equation} 

\noindent where the $Pe_{n,N}^{g,h}(\cos(2x))$ are the exceptional Jacobi orthogonal polynomials \cite{sasaki} 

\begin{eqnarray}
Pe_{n,N}^{g,h}(\cos(2x)) & = & a_{n,N}^{g,h}(\cos(2x)) P_n^{g+N-\frac{1}{2},h+N-\frac{1}{2}}(\cos(2x)) + \\
& &  b_{n,N}^{g,h}(\cos(2x)) P_{n-1}^{g+N-\frac{1}{2},h+N-\frac{1}{2}} (\cos(2x)), \nonumber
\end{eqnarray}

$$
a_{n,N}^{g,h}(x) \! =  P_N^{-g-N-\frac{3}{2},h+N-\frac{1}{2}}(x)  + 
$$
$$
+ \frac{2n(h \! +\! N \! - \! g \! - \! 1 \!) P_{N-1}^{-g-N+\frac{1}{2},h+N-\frac{1}{2}}(x)}{(h \! + \! 2N \! - \!2 \! - \!g)(g \! + \! h \! + \! 2n \! + \! 2N \! -1)} - 
\frac{n(2h \! + \! 4N \! - \! 3)P_{N-2}^{-g-N+\frac{1}{2},h+N-\frac{1}{2}}(x)}{(2g \! + \! 2n \! + \! 1)(h \! + \! 2N \! - \!  g \! - \! 2)} ,
$$

$$
b_{n,N}^{g,h}(x) = \frac{(h+N-g-1)(2g+2n+2N-1)}{(2g+2n+1)(g+h+2n+2N-1)} P_{N-1}^{-g-N+\frac{1}{2},h+N-\frac{1}{2}} (x).
$$

\subsection{ { \bf The stationary hydrogen atom}}

Finally, we consider the case of the hydrogen atom. Let us consider as prepotential the following 

\begin{equation}
W = -\frac{\mu}{2(N-l)}x - l \ln x +\ln (L_N^{-2l-1}(\frac{\mu x}{N-l})), 
\end{equation}

\noindent where $L_n^l(x)$ are the Laguerre polynomials. We replace this prepotential $W$ in equations (\ref{susyzero}) and (\ref{susy})

\begin{equation}
\begin{cases}
a = -\partial_x + W_x ,\\
a^{\dagger} = \partial_x + W_x,
\end{cases}  
\end{equation}

\noindent such that equation (\ref{prepoteq})  takes the form
$$W_x^2 + W_{xx} = \sqrt{2k} + 2 V_q = \frac{\mu^2}{4(l-N)^2} - \frac{\mu}{x} + \frac{l(l+1)}{x^2},$$

\begin{equation}
\begin{cases}
a^\dagger a \rho = 0, \rho=e^W , \\
(a a^\dagger - 2 \sqrt{2k_1} ) \psi = 0,
\end{cases}
\end{equation}

\noindent where the wavefunction $\psi$ and the parameter $k_1$ are given by 

\begin{equation}
\psi = a ( e^{-\frac{\mu x}{2(l+n+1)}} x^{l+1} L_n^{2l+1} (\frac{ \mu x}{l+n+1}) ), \quad k_1 = \frac{2 \mu^4 (N^2 - (n+1)^2 -2l(N+n+1))^2}{4^4( l-N)^4 (l+n+1)^4}.
\end{equation}

\noindent In particular if we set $n=N=0$ we obtain an explicit form for the function $b_1$

\begin{equation}
b_1 = e^W \psi = e^{\frac{\mu x}{2l(l+1)}} \frac{2l+1}{2l} (2l - \frac{x \mu}{l+1}),
\end{equation}

\noindent from which we get for the classical potential $V_c$

\begin{equation}
V_c = k _1 \frac{b^2}{b_u^2} = \frac{(2l+1)^2}{8l(l+1)} \left( \frac{l(l+1)}{u^2} - \frac{\mu}{u} + \frac{\mu^2}{4 l (l+1)} \right). 
\end{equation} 

\noindent As already shown for the quantization of   (\ref{schroedingerPIV}), (\ref{VPIV}) and   (\ref{schroedingerPV}), (\ref{VPV}) we verify that if we chose $b_1(\cdot) = b_2(\cdot)$ (which is the choice adopted in \cite{zabrodinzotov} in order to have $b|_{x=u} = 0$) then we obtain a classic potential $V_c(u)$ whose limit for large $l$ coincides with $V_q(u)$. However, we should remark that in this case the energy of the system turns out to be fixed and corresponds to that of a particle moving in a Kepler/Coulomb potential on a circular orbit.

\section{The Harmonic oscillator and the exact solution of its LSP}
In the previous sections we have shown that any one-dimensional Newton equation (\ref{newton1}) can be associated with a LSP which coincides with a time-dependent Schroedinger equation. The aim of this section is to discuss thoroughly an explicit example with the goal of providing the exact solution of the LSP (which in general is a superposition of solutions for the time-independent Schroedinger equation) for any given solution of its classical counterpart (\ref{newtonharmonic}).  

\noindent Let us consider the Harmonic oscillator potential obtained from the solution of (\ref{vc}) with  $b = x^2$ and the constant $k_1 = 2 \omega^2,k_2 =-4E, k_3 = -2 l^2$ . This choice produces the following Lax pair (\ref{start}) with potential matrices of the form

\begin{equation}
\label{uharmonic}
\bf{U} = \left( \begin{array}{cc} 0 & x^2 - u(t)^2 \\ \frac{\frac{l^2 - \frac{1}{4}}{x^2} + \omega^2 x^2 -\frac{l^2}{u(t)^2}-\omega^2 u(t)^2 - \dot{u}(t)^2- \frac{1 - 2iu(t) \dot{u}(t) }{x^2 - u(t)^2 }}{x^2 - u(t)^2} & 0 \end{array} \right) ,
\end{equation}

\begin{equation}
\label{vharmonic}
\bf{V} = \left( \begin{array}{cc} \frac{-i -2 u(t) \dot{u}}{2(x^2 -u(t)^2)} & ix \\ \frac{4l^2x^2+u(t)^2 (1-4l^2-4x^4 \omega^2 + 4 x^2 (\omega^2 u(t)^2 + \dot{u}(t)^2))}{4xi u(t)^2 (x^2 -u(t)^2 )^2} & \frac{i +2 u(t) \dot{u}}{2(x^2 -u(t)^2)} \end{array}\right).
\end{equation}

\noindent It is possible to verify by direct calculation that the ZCC is equivalent to the Newton equation (\ref{newton1}) for a classical particle moving under the oscillator potential plus a centrifugal barrier

\begin{equation}
\label{newtonharmonic}
\partial_t {\bf U} - \partial_x {\bf V} + [{\bf U} , {\bf V}] = 0 \equiv \ddot{u} - \frac{l^2}{u(t)^3} + \omega^2 u(t)  = 0,
\end{equation}

\noindent which is satisfied for the function

\begin{equation}
u(t) = \sqrt{\frac{E}{\omega^2} - \frac{\sqrt{E^2 - l^2 \omega^2}}{\omega^2} \sin (2 \omega t)}.
\end{equation}

\noindent The LSP (\ref{eq9lsp}) for the Lax pair (\ref{uharmonic}) and (\ref{vharmonic}) reduces to the following system of two linear PDEs

\begin{eqnarray}
\label{lspharmonic0}
 i {\phi_1}_t + E \phi_1 = -\frac{1}{2} {\phi_1}_{xx} + \left( \frac{l^2-\frac{1}{4}}{2x^2} + \frac{\omega^2}{2} x^2 \right) \phi_1 , \\ 
\label{lspharmonic} 2i (x^2 - u(t)^2 ) {\phi_1}_t = (1 -2 i \dot{u}(t) u(t) ) \phi_1 -2x {\phi_1}_x ,
\\ \phi_2 = \frac{{\phi_1}_x}{x^2 - u^2(t)},
\end{eqnarray}

\noindent where the functions $\phi_1$ and $\phi_2$ are the two component of the wave function vector 

\begin{equation}
\boldsymbol{\Phi} = \left( \begin{array}{c} \phi_1 \\ \phi_2 \end{array} \right) 
\end{equation}

\noindent Let us expand the function $\phi_1$ as a series of stationary solutions of the Schroedinger equation (\ref{lspharmonic0})

\begin{equation}
\label{seriessolution}
\phi_1 = \sum_n c_n e^{-i \epsilon_n t} \chi_n^l , \quad \epsilon_n = \omega (2n+l+1) - E, \quad \quad \chi_n^l \equiv e^{-\frac{\omega x^2}{2}}x^{l+\frac{1}{2}} L_n^l (\omega x^2) .
\end{equation}

\noindent Replacing (\ref{seriessolution}) in (\ref{lspharmonic}) and taking into account the following relations

\begin{eqnarray}
x \partial_x \chi_n^l & = & (n+1) \chi_{n+1}^l -\frac{1}{2} \chi_n^l - (l+n) \chi_{n-1}^l , \\
-\omega x^2 \chi_n^l & = & (n+1) \chi_{n+1}^l - (2n + l +1 ) \chi_n^l + (n+l) \chi_{n-1}^l ,\\
u^2 \partial_t e^{-i \epsilon_n t} & = & \frac{-i E \epsilon_n}{\omega^2} e^{-i \epsilon_n t} \! + \! \frac{\epsilon_n \sqrt{E^2 \! - \! l^2 \omega^2}}{2 \omega^2} (e^{- i \epsilon_{n-1} t } \! - \!  e^{-i \epsilon_{n+1}t}) , \\
-2 u \dot{u} e^{-i \epsilon_n \! t} \!\! & = & \frac{\sqrt{E^2 - l^2 \omega^2}}{\omega} (e^{- i \epsilon_{n \! - \!1} \! t } + e^{-i \epsilon_{n \! + \!1} \! t}) ,
\end{eqnarray}

\noindent we arrive at the following set of equations determining the coefficients $c_n$ of the series (\ref{seriessolution})

\begin{equation}
e^{-i \epsilon_n t} \left( \alpha_n \chi_{n+1}^l + \beta_n \chi_n^l + \gamma_n \chi_{n-1}^l \right) = 0 , \quad \forall n  
\end{equation}

\begin{eqnarray}
\alpha_n & = & i c_{n+1} \frac{\sqrt{E^2-l^2 \omega^2}}{\omega} \left( 1 + \frac{\epsilon_{n+1}}{\omega} \right) + 2 c_n (n+1) \left( \frac{\epsilon_n}{\omega} -1 \right) = 0 , \\
\beta_n & = & 2 c_n \left( 1 - \frac{\epsilon_n}{\omega}(2n+l+1) + \frac{E \epsilon_n}{\omega^2} \right) = 0, \\
\gamma_n & = &  2 c_n (n+l) \left( \frac{\epsilon_n}{\omega} + 1 \right) + i c_{n-1} \frac{\sqrt{E^2 -l^2 \omega^2}}{\omega} \left( 1 - \frac{\epsilon_{n-1}}{\omega}\right) = 0.
\end{eqnarray}

\noindent This system of equations can be solved for $c_n$ with a series of two terms if we set the classical energy $E$ to the value $E = \omega(2n+l+2)$ . With this choice we determine the coefficients $c_n$ to be

\begin{equation}
\frac{c_{n+1}}{c_n} = \frac{-i \sqrt{n+1}}{\sqrt{n+l+1}} , \quad c_i=0  \quad \!\! \{i \neq n,n+1\}, \quad n \geq 0 .
\end{equation}  

\noindent Therefore a classical particle moving with energy $E = \omega (2n + l + 2), n \geq 0$ is associated to a $\phi_1$ which is a superposition of two quantum states with energies $E_1 = \omega (2n + l + 1)$ and $E_2 = \omega (2n + l + 3)$ 

\begin{equation}
\phi_1 = \frac{1}{<\phi_1 | \phi_1 >} \left( e^{-i \omega t}e^{-\frac{\omega x^2}{2}} x^{l+\frac{1}{2}} L_n^l (\omega x^2) - \frac{i \sqrt{n+1}}{\sqrt{n+l+1}} e^{i \omega t}e^{-\frac{\omega x^2}{2}} x^{l + \frac{1}{2}}L_{n+1}^l (\omega x^2) \right) , 
\end{equation}

\noindent which produces a probability $<\phi_1 | \phi_1 >$ which oscillates with frequency $\omega$.

To conclude let us consider the case of a stationary particle of energy $E = \omega l$. In this case the solution for $\phi_1$ turns out to be 

\begin{equation}
\phi_1 = \frac{e^{-i \omega t} e^{-\frac{\omega x^2}{2}} x^{l+\frac{1}{2}}}{<\phi_1 | \phi_1 >} ,
\end{equation} 

\noindent whose probability, as expected, turns out to be time-independent.

\section{Concluding remarks and future outlook}
The main result of the paper is the application of the quantization procedure in the sense of the SZZ  \cite{suleimanov} \cite{zabrodinzotov} to any time-independent potential. We have shown that such a quantization can be realized up to the solution of the Gambier XXVII equation. In particular it is shown that explicit solutions can always be computed for any shape-invariant potential.
Particular examples have been analyzed for classical and exceptional orthogonal Laguerre and Jacobi polyomials. Finally the solution of the LSP associated with the quantization of the Harmonic oscillator is provided explicitly. The classical energy $E = \omega (2n + l +2)$ turns out to be the mean value of the energy eigenvalues of the two wavefunctions in the series (\ref{seriessolution}) which satisfy the LSP $E_1 = \omega (2n + l +1) < E < E_2 = \omega (2n + l + 3)$ establishing in this way a new connection between the classical Newton equation (\ref{newtonharmonic}) and its quantum counterpart (\ref{lspharmonic}). There are reasons to expect that this connection can also be found for more general quantization procedures with time-dependent potentials. An analysis of equations (\ref{vc}) and (\ref{vq}) for potentials similar to the one studied in sections 2 and 3 can provide us with an explicit form of the wavefunction satisfying the LSP (\ref{eq2}). Since the ODE (\ref{vq}) constitutes a special case for any time-independent potential, it is evident that our approach can be applied to systems which describe much more diverse types of potentials.  Another interesting avenue for future research could include the study of soliton surfaces based on the LSP for quantum Hamiltonian systems. These surfaces are directly expressed in terms of the wavefunction $\Phi$ satisfying the associated LSP (\ref{start}) of the considered model (see e.g. \cite{GPR} \cite{fokas}). A visual image of such surfaces reflecting the behaviour of solutions can be of interest, providing information about the properties of these surfaces, which otherwise would be hidden in some implicit mathematical expression. These tasks will be undertaken in a future work. 

 \section{Acknowledgements}
A.M.G.'s work was supported by a research grant from the Natural Sciences and Engineering Council of Canada (NSERC). D.R. wishes to acknowledge a fellowship from the Laboratory of Mathematical Physics of the Centre de Recherches Math\'ematiques CRM (Universit\'e de Montr\'eal).    

\end{document}